# Recommendations on Future Operational Environments' Command Control and Cyber Security

K. Goztepe

*Abstract*—It is a well-known fact that today a nation's telecommunication networks, critical infrastructure, and information systems are vulnerable to growing number of attacks in cyberspace. Cyber space contains very different problems involving various sets of threats, targets and costs. Cyber security is not only problem of banking, communication or transportation. It also threatens core systems of army as command control. Some significant recommendations on command control (C2) and cyber security have been suggested for army computing environment in this paper. This study addresses priorities of "what should be done for a better army cyber future" to cyber security researchers.

## I. Introduction

INITIALY, wars were conducted via physical instruments and signs like whistles or flags. Especially with the usage of telegraph in Crimean War [1] in the 19th century, remote command and control (C2) became possible. The rapid development of the instruments like satellites [2] within the space dimension increased the range of C2, while the improvement of ground positioning systems [3] increased the precision of C2, and lastly computers and internet increased the obscurity of C2.[4]

Although the C2 instruments have changed through the history [5], its crucial role in warfare has kept its importance; it is a necessity to have sheltered, mobile, and modern C2 centers, while it is indispensable to train the leaders to effectively put to use this capability [6].

The more electronics and mobile devices are improved the more the air, land, sea, and space become cyber environments [7]. The cyber security became a state issue, where a cyber-threat for a security system of a state is accepted as a casus belli. Cyber-attack has been defined as "an attack, via cyberspace, targeting an enterprise's use of cyberspace for the purpose of disrupting, disabling, destroying, or maliciously controlling a computing environment/infrastructure; or destroying the integrity of the data or stealing controlled information" by National Institute of Standards and Technology (NIST) [8]. Besides the cyber defense and cyber-attack capability has also gained importance; however, there is no sufficient legal basis for such an operation [9]. That is why, states focus on identifying the imminent threat and staving it off by state-sponsored (non-state) cyber-attacks, instead of directly applying it.

## II. What Should Be Done for a Better Cyber Future?

### A. Purpose of the Study

Complexity of modern cyber security is increasing all over the world. Today it is a fact that manage cyber threats and cyber incidents without leveraging various collaboration instruments with different partners is not possible. Cyber-attacks constitute a serious challenge to national security and demand greater attention from all public and civil establishments. This study mention main topics about cyber security especially for army. The main issues needed to be done in terms of future operational environments' command control and cyber security are given below.

### B. Main Issues for Cyber Security

Interoperability is a fundamental principle for the operation of joint and coalition forces (Fig.1). Army should have the capability to operate in any place of the world with the ability of "interoperability for joint and coalition operations". It is also crucial that the joint situational picture of the operational area be presented to joint headquarters. Additionally, cross-functional structure among the systems should be formed, and centralized C2 systems should be transformed to decentralized C2 systems.

It is a necessity to introduce programs about cyber security in the universities and to encourage the education of qualified individuals about it. The purpose of the cyber exercises held by armies or scientific centers, should be to create awareness. The situational awareness will not be sufficient in future's sophisticated and ambiguous environment. In order to provide all of the functional areas in a C2 headquarters work jointly, situational awareness should be improved by detecting the

Kerim Goztepe, IE, Ph.D. War Colleges Command, Army War College, Dept. of Operations & Intelligence, Yenilevent, 34330, Pbx: +90 212 398-0100/3262, İstanbul-Turkey,  e-mail: kerimgoztepe@gmail.com.

factors in the environment, apprehending the current situation, and envisaging the future situation (tendency awareness).

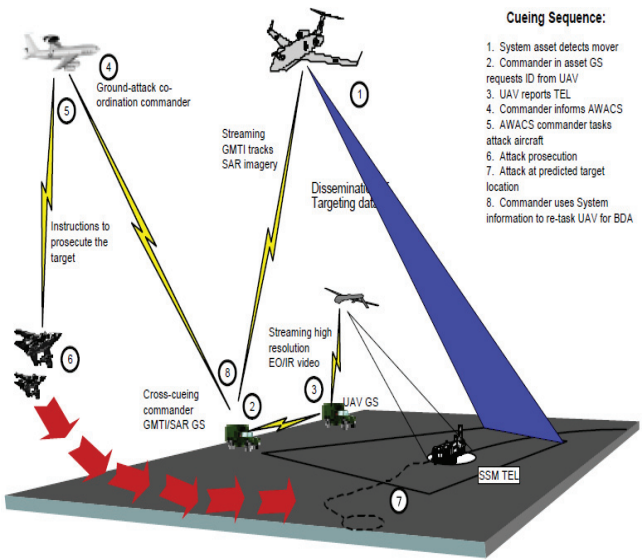

Fig.1. Interoperability [10]

The armies will increase their Network Enabled Capabilities (NEC) in order to eradicate the obscurity of the chaotic future warfare environments, and to provide right information, in the right time, and in the right place (Fig.2) [10]. However, the more the NEC improved, the more the systems/platforms will be exposed to cyber-attacks. 'Embedded Cyber Security Technique' and 'Operational Area Identification and Introduction Systems' [11] should be used to restrain from those attacks.

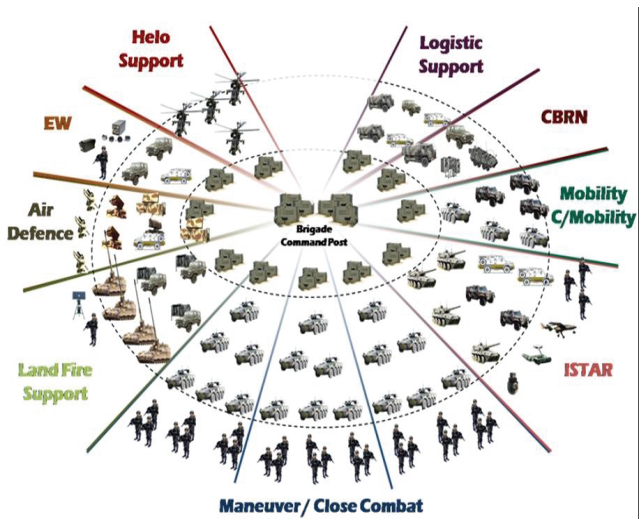

Fig. 2. Army network enabled capability [10]

It is of high importance that an Army, which has mission functions that entails different C2 structures, must have the capability to conduct via a single and centralized C2 system. In addition to this, special weaponry systems for army should be compatible for a sophisticated C2. Today the C2 systems in army, depends on software and hardware, which are open to cyber-attacks. Thus, 'National Tactic Data Link' is necessary for securing the systems.

Internet search engines are the most sophisticated spy software of the world, since they store all of the data about the users and draw the data map of the world [12]. They can provide the educational levels of the countries, the contacts of the individuals, their social circles, and expenses (Fig.3). This capability renders those engines as the indispensable environment for intelligence that is able to analyze so much data. In order to preserve from the threats of that capability, it is vital to use complicated passwords and to use these systems quantum sufficit.

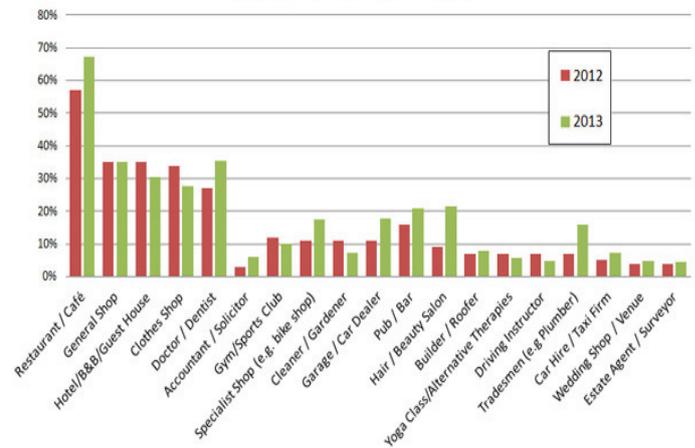

Fig. 3. Search engine usage sample [13]

Depending on the last developments on Cryptology, authentication, access control (for 2 or more people), authorization, strong authentication mechanisms, fast and low-source-spending crypto algorithms, asymmetric crypto algorithms resistant to quantum computers (rather than symmetric crypto) should be used in order to provide data security of the NEC systems. It is known that trying to secure all kinds of data may decrease the effectiveness of the systems or put systems out of action. Thus, it is essential that, which data is worth security, be decided effectively. Instead of commencing the crisis and operations process with intelligence function and following with operation function, the process should be defined from the crisis phase cross functionally and jointly.

Network enabled capability gains the upper hand. In order to make use of this capability, it is vital to have communications networks that enable current systems, platforms, and nodal points function together. Additionally, information and nodal systems are required to process the data and prepare it compatible with the users' needs. Moreover, an organization structure and a training system (of the users and the decision makers) that effectively profits from the network-enabled capability are indispensable.

*C. Command Control and Tactical Issues*

In the near future, robotic armies and unmanned land vehicles will be under command, and most of the platforms/systems will depend on satellites. In order to remove weaknesses stemming from satellite signal intervention, a 'National GPS System' is vital. By using national GPS system Decision Support Systems may provide alternative solutions about deployment of artillery weapons and weapon systems, ammunition effect analysis, desired effect, the sort and amount of the ammunition. Metro techniques affect hit rate % 67 percent in the long-range. In this regard, service-based metro systems will be widespread. It is possible to take better decisions than the previous ones, and artificial intelligence systems may be helpful for understanding the intention/course of action of the enemy using Learning Systems.

'Vetronics' systems were introduced with the developments in the land vehicles (Fig.4) [14],[15]. In the middle of the battle, if the commander needs to change his vehicle due to mechanic problems, he will have the capability to use the new vehicle as a command and control vehicle with the help of 'mountable systems' [16].

With the help of tactic wireless net communication system applications, all of the forces, particularly Navy, will have the capability to communicate worldwide securely, based on completely national means.

Studies and researches are in progress on the use of 'Service Based Architecture' by the mobile elements on the field and on the tactical field communication nets. It is predictable that service based architecture will be used on a broader band in between command centers over brigade level in the short and medium term. The C2 systems situational awareness will be available for the single soldier on the field as well as the high level command, parallel to the technologic developments, and functional capabilities will increase with that. Identification system is as important as detection system on the unmanned aerial vehicles, in that context, by help of Terahertz technology, which follows the terahertz rays that objects radiate, the identification of the weapons, explosive, chemical, and biological materials and similar security inspections will be available.

It is possible to use network enabled capability in strategic, operative, and tactic levels; however, it will be difficult to use it in single soldier / single vehicle level regarding the hardships of the combat. Thus, it is vital to define the levels that network enabled capabilities are planned to operate.

With the help of a data exchange system, different platforms should have the ability to activate the weapon systems of each other. Most of the systems in army are compatible with platform-based operation, rather than network enabled ones, which may cause problems in the long run.

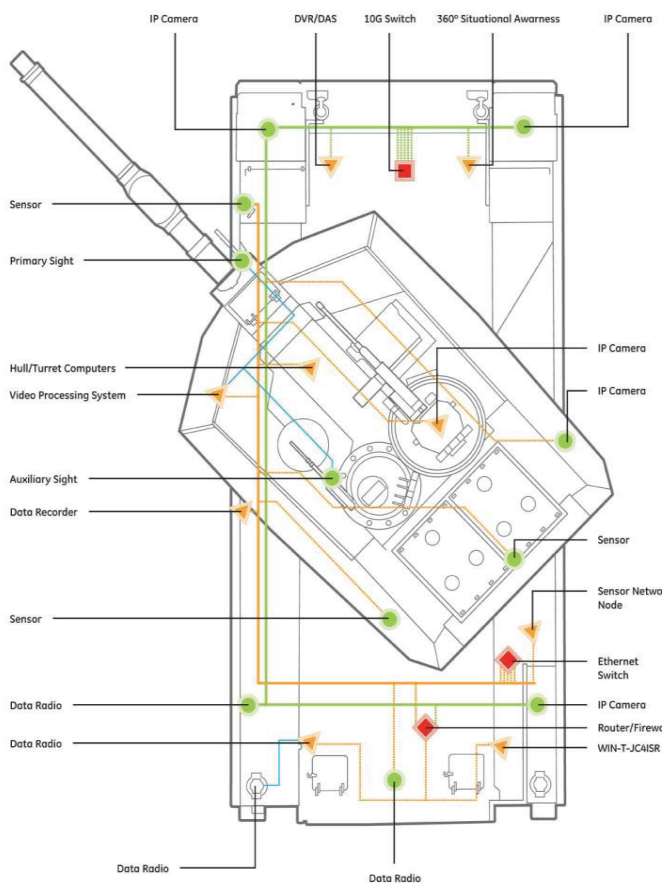

Fig. 4. Vetronics for a main battle tank [17]

## III. CONCLUSION

Recommendations on future operational environments' command control and cyber security has been discussed in this paper. The studies' goal is to make curious researchers some main topics about army and cyber security issues.

It is known that in order to form a new effective defense technology base, focused on the needs of army, long term planning, application, and tracking activities should be handled with strategic management approach [18],[19]. In this context, regarding command, control and information technologies excellence nets C4ISR technologies, which are compatible with network centric capability, should be prioritized. Autonomous command and control technologies in land, navy and air force vehicles, real time data integration and data fusion, cyber defense, strategy and tactics improvement, protected core, and national nets should also be prioritized.

It is fundamental that national software and hardware be used in order to be secure from or least affected from threats. Human-education-technology synergy be provided for a sufficient cyber defense, and that a national cyber army and a cyber-defense supreme board be formed. It is also important to promote a cyber-market while improving the research & development about cyber security.

One of the fundamental concept in battlefield environment is information flow and security of obtained information. It is well known fact by cyber experts that physical components added to a cyber-system considerably increase the difficulty of determining information flow.

Todays' armies have sophisticated command control infrastructures includes complex interactions of cyber components. Confidentiality of operational environment requires advanced security integration models. I believe that this work provides an approach for some of the complexities involved in command control and cyber security protection for army.

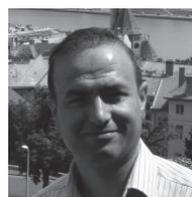

**Dr. Kerim Goztepe** received the B.S.degree in System Engineering from Turkish Military Academy in 1998. He was accepted in Marmara University and received M.S. degree in Engineering Management in 2003. He started doctoral education in Industrial Engineering, Institute of Science and Technology at Sakarya University and graduated in 2010. He is interested in fuzzy logic, neural network, multicriteria decision making and cyber security for open source systems. Dr. Kerim Goztepe is an author of more than thirty refereed papers, and editor-in-chief of "Journal of Military and Information Science".